\documentclass[electronic]{vgtc}             

\graphicspath{{figures/}{pictures/}{images/}{./}} 

\usepackage{times}                     

\usepackage{tabu}                      
\usepackage{booktabs}                  
\usepackage{lipsum}                    
\usepackage{mwe}                       

\usepackage{mathptmx}                  
\usepackage{amsmath}
\usepackage{amsfonts}

\onlineid{0}

\vgtccategory{Research}

\vgtcinsertpkg

\title{Learning Martial Arts Motion Skills for Animating Opponents \\in VR Sparring Games}

\author{Jiqing Wen\thanks{e-mail: jwen31@asu.edu}\\ %
        \scriptsize Arizona State University %
\and Qixuan Shi\thanks{e-mail: qshi36@asu.edu}\\ %
     \scriptsize Arizona State University %
\and Robert LiKamWa\thanks{e-mail: likamwa@rice.edu}\\ %
     \scriptsize \centering Rice University %
\and Pavan Turaga\thanks{Pavan.Turaga@asu.edu}\\ %
     \scriptsize Arizona State University }

\abstract{
Animating virtual opponents is an important yet challenging task in game development. In martial arts games, opponent motions must not only be physically plausible but also exhibit appropriate technique, timing, transitions, and target-directed intent. Traditional game animations often rely on handcrafted clips, state machine transitions, and procedural adjustments, which can be labor-intensive and may limit the behavioral diversity of non-player characters. This paper presents a martial arts motion skills learning framework for character animation in virtual sparring games. Inspired by Tang Soo Do, we collected a martial arts motion capture dataset containing isolated attacking and defending techniques, technique combinations and transitions, and contextual sparring variations. We first pretrained a reusable repertoire of martial arts motion skills from the captured data using reinforcement learning (RL) with adversarial style rewards, and then adapted these skills with task-specific rewards for target-directed striking and threat-aware evasion. We evaluated the motions generated by the learned policy through a user study, in which participants assessed their physical plausibility, smoothness, and martial arts form. To demonstrate their applicability to interactive gameplay, we developed a VR sparring prototype for Meta Quest headsets. The prototype combines policy-generated motion clips with lightweight state-based control to support interactive opponent behavior. Together, these findings support the feasibility of using a learned martial arts skill repertoire as a motion foundation for animating opponents in VR sparring games.
} 


\begin{document}

\maketitle

\section{Introduction}
\begin{figure*}
  \centering
  \includegraphics[width=\linewidth]{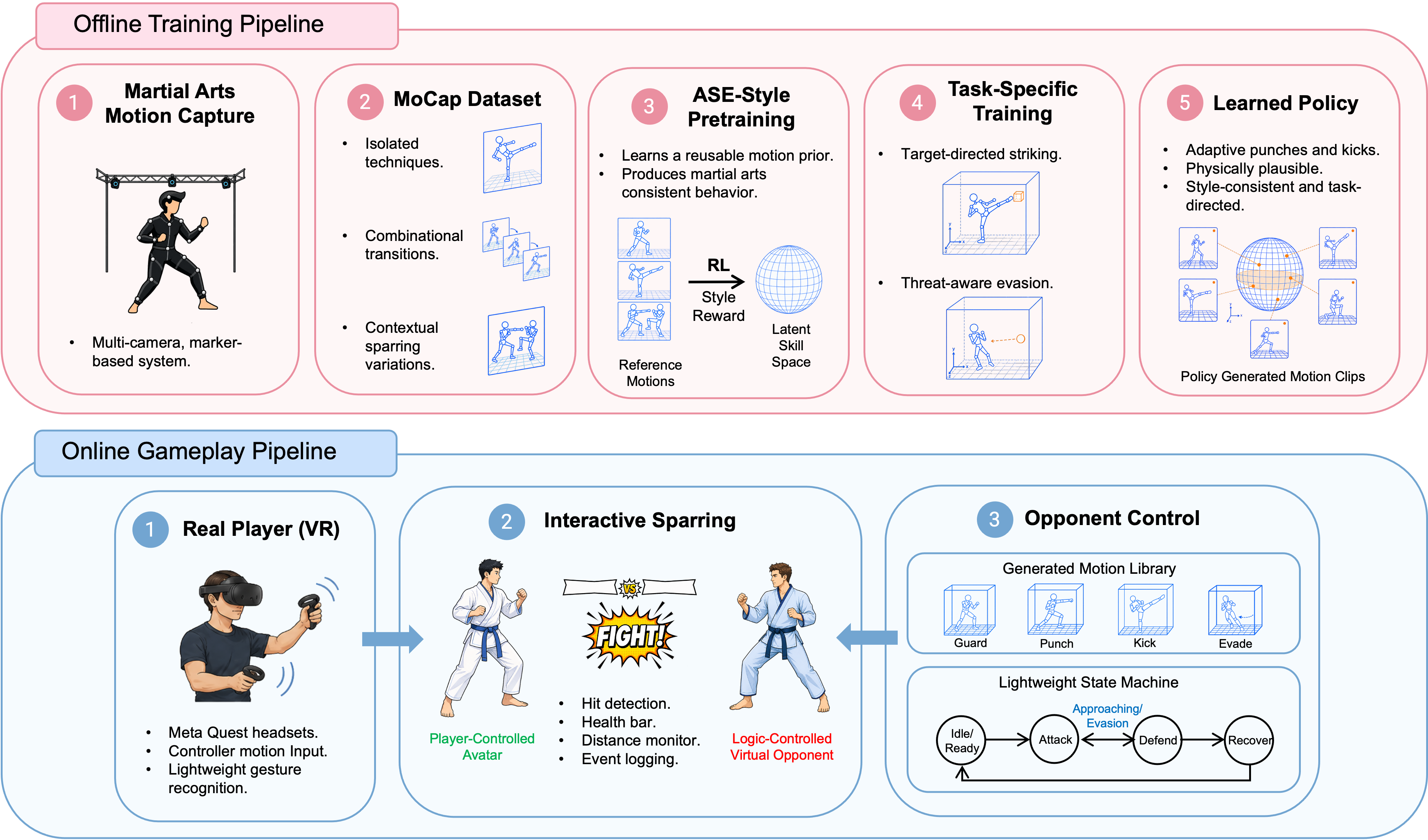}
  \caption{Overview of the Virtual Martial Arts Gaming System.}
  \label{fig:systemdiagram}
\end{figure*}
Traditional animation production involves multiple stages, including character design, 3D modeling, rigging, and keyframe-based motion authoring. While modern game engines such as Unity and Unreal provide animation systems to manage logic and state transitions, most game animations are still manually scripted and require substantial human effort. Moreover, unlike film animation with fixed timelines, interactive applications demand animations to adapt dynamically to real-time events, further increasing development complexity and cost. Direct mocap manipulation is also not enough, because it can preserve capture noise, marker errors, or self-collision caused by the differences in body proportions between human performers and virtual characters.

Deep learning-based animation synthesis has greatly improved real-time character animation by replacing hand-authored state graphs with data-driven or neural motion generators \cite{starke2020local, starke2021neural, starke2022deepphase}. However, many of these methods are still primarily kinematic: they synthesize visually plausible motion, but they do not necessarily guarantee physical interaction, task success, balance under contact, or adaptive opponent behavior.

Recent advances in robotics and physics-based humanoid control provide a promising direction for addressing these limitations. Instead of treating character animation only as pose generation, physics-based learning formulates motion as a control problem, where an embodied agent must generate actions that satisfy both stylistic and task-driven objectives. Generative motion models further expand this possibility by reducing repetitive manual authoring effort while increasing the range of behaviors that virtual characters can produce. More importantly, learned motion policies can generate adaptive, varied, and less predictable behaviors through motion priors and emergent motion combinations, rather than relying solely on fixed animation sequences. Beyond increasing behavioral diversity, task-specific training examines whether the learned repertoire can be adapted to functional combat objectives, such as striking a target or evading an incoming threat. Together, these capabilities can support more realistic and varied opponent behaviors, with the potential to provide human players with less predictable and more challenging sparring experiences.

Motivated by these advances, we propose a skill-based learning framework for generating physically plausible, martial arts-specific character motions. Figure~\ref{fig:systemdiagram} presents the broader vision of a virtual martial arts game that connects offline motion learning with online interactive gameplay. Within this vision, the present work focuses on the offline training pipeline, which uses martial arts demonstrations to learn a reusable repertoire of movements and subsequently adapts the learned repertoire for target-directed striking and threat-aware evasion. We evaluate the resulting motions through task performance measures and a user study of their perceived motion quality. To explore their potential for interactive gameplay, we retarget motions generated by the learned policies as animation clips and integrate them into a lightweight Unity sparring prototype. Predefined movement thresholds recognize the player’s controller gestures and determine the requested action state, after which a motion is selected from the skill library associated with that state. The player-controlled character spars with a virtual opponent fully governed by the game logic and the retargeted clips. The player receives game feedback such as distance visualization, health bars, sound effects, and event logs. The prototype therefore serves as a proof of concept for integrating the learned motions into virtual sparring; executing the learned policies online to support physics-based interaction remains future work.

Learning-based control for humanoid robots often builds on imitation learning (IL) \cite{xie2020deep, seo2023deep}, where the goal is to map states to actions based on expert demonstrations. Behavioral Cloning (BC) provides a straightforward approach to this by treating the problem as supervised learning; however, it often suffers from distributional shift when the robot encounters unseen states. To overcome these limitations, researchers have turned to RL combined with generative models to create more robust controllers. For instance, the pioneering work AMP \cite{peng2021amp} employs a Generative Adversarial Network (GAN) to produce style rewards, allowing the agent to bridge the gap between rigid imitation and flexible RL by learning from motion priors. Building upon this, ASE \cite{peng2022ase} scales the approach by encoding vast datasets into a latent space, which can then be sampled to synthesize complex behaviors. In this study, we utilized Tang Soo Do as a representative model for martial arts movement. We incorporated the Tang Soo Do training progression and technical hierarchy into our humanoid training paradigm and the gaming system's scoring mechanism. We improved the training flow by decomposing the complex combat behavior into two foundational subtasks: 1. target-directed striking, in which the humanoid learns to strike virtual targets using punches and kicks; and 2. threat-aware evasion, in which the humanoid learns defensive movements to avoid incoming threats. In this way, we can produce more realistic martial arts motions.

Our contributions include the following:
\begin{enumerate}
    \item We collected a Tang Soo Do-inspired motion capture dataset using a structured three-phase protocol that includes isolated techniques, short combinations and transitions, and contextual sparring variations.
    \item We developed a staged ASE-based training pipeline that combines adversarial motion pretraining with task-specific adaptation, enabling physics-based humanoid agents to perform target-directed strikes and threat-aware evasions while retaining the learned martial arts motion characteristics.
    \item We evaluated the learned motions through a practitioner user study and demonstrated their applicability in a lightweight Unity prototype, showing their potential for physically plausible character animation that is recognizable as martial arts motion in virtual sparring games.
\end{enumerate}
\section{Related Works}
We review three areas related to our system: character animation synthesis, humanoid behavior learning, and interactive agents in VR combat games. Character animation synthesis generates visually plausible movements from motion data or generative models, often without explicit physics; humanoid behavior learning trains simulated characters to perform motor skills under balance, contact, and task constraints; and game-agent research examines how characters perceive player actions, select responses, and plan during interaction. Together, these areas inform our approach to learning physics-based martial arts skills for interactive virtual sparring.

\subsection{Character Animation Synthesis}
Classical data-driven animation methods reduce the authoring burden of handcrafted clips and scripted transitions by reusing and recombining motion-capture data. Early motion interpolation and blending methods synthesize new motions by combining similar examples according to control parameters such as speed, direction, style, or target position \cite{wiley1997interpolation, rose1998verbs}. Beyond interpolation, motion graphs sequence compatible captured segments \cite{Kovar2002MotionGraphs}, whereas parameterized motion spaces blend registered examples into controllable variations \cite{kovar2004automated}. Interactive control can also be achieved through runtime database search in motion matching \cite{clavet2016motion} or continuous controllers learned over mocap-based motion spaces \cite{treuille2007near}.

While these classical methods support user control to varying degrees, their flexibility is still constrained by predefined parameters, graph connectivity, or database coverage. Deep learning-based methods address this limitation by learning motion representations directly from data, allowing character motion to be generated from high-level controls, character states, scene context, or user commands. Holden et al. \cite{holden2016deep} introduced a neural framework for motion synthesis and editing on a learned motion manifold, and later proposed Phase-Functioned Neural Networks (PFNN), where network weights are modulated by motion phase to generate responsive locomotion from user control, previous character state, and scene geometry \cite{holden2017phase}. Subsequent work further improved neural character control for complex and contact-rich motions. Starke et al. proposed neural frameworks for character-scene interaction, local motion phases, and martial arts motion synthesis, showing that learned representations can support asynchronous body timing, multi-contact movements, and controllable combinations of diverse motion skills from raw mocap data \cite{starke2021neural, starke2020local, starke2022deepphase}. These methods mark a shift from explicit clip search and hand-authored transition logic toward learned motion models that can synthesize smooth and responsive animation under continuous control.

More recently, generative models have further expanded the scope of motion synthesis. Probabilistic latent-variable models can generate diverse character animations under weak control signals \cite{ghorbani2020probabilistic}, while GAN-based approaches such as GANimator synthesize novel motion from limited motion examples \cite{li2022ganimator}. Diffusion-based models have also enabled text-driven human motion generation; for example, MotionDiffuse \cite{zhang2024motiondiffuse} generates human motion sequences conditioned on natural-language descriptions, supporting probabilistic text-to-motion mapping and diverse synthesis. Recent systems such as MotionBricks \cite{wang2026motionbricks} and Kimodo \cite{rempe2026kimodo} suggest a broader trend toward large-scale controllable animation models, where motion is generated from learned representations under user-specified constraints rather than selected from a small set of authored clips. However, most of these approaches remain primarily kinematic: they generate visually plausible motion trajectories, but do not directly guarantee physical balance, contact forces, impact response, or task success. 

\subsection{Humanoid Behavior Learning}
Physics-based humanoid control commonly combines RL with motion imitation to reproduce reference motions while maintaining balance and satisfying contact and other dynamic constraints. DeepMimic \cite{peng2018deepmimic} was a landmark example, showing that RL with carefully designed pose, velocity, end-effector, and root tracking rewards can produce robust physics-based imitation of diverse skills, including locomotion, acrobatics, and martial arts. Subsequent methods improved the scalability and flexibility of this paradigm. Residual Force Control introduced auxiliary residual forces to compensate for dynamics mismatch when imitating highly agile motions \cite{yuan2020residual}, while adversarial approaches such as AMP \cite{peng2021amp} replaced hand-designed tracking rewards with motion priors learned from unstructured motion clips. ASE \cite{peng2022ase} further extended this idea by learning reusable adversarial skill embeddings, enabling high-level tasks to be solved through a learned latent motor space rather than training a separate low-level controller from scratch.

Recent work has increasingly focused on scaling physics-based humanoid control from individual skills toward general-purpose motion representations. Methods such as CALM and C$\cdot$ASE learn controllable latent spaces that allow virtual characters to generate diverse and directable behaviors from large motion datasets \cite{tessler2023calm,dou2023c}. PHC scales motion imitation to large unstructured motion collections and improves fault tolerance under noisy references and unexpected falls \cite{luo2023perpetual}. PULSE further distills broad motion imitation capabilities into a universal humanoid motion representation that can be reused for downstream tasks such as striking, terrain traversal, and VR tracking \cite{luo2024universal}. More recent systems such as MaskedMimic and PARC explore unified or iterative frameworks for synthesizing physics-based character motion from flexible constraints, partial motion descriptions, or augmented motion data \cite{tessler2024maskedmimic,xu2025parc}. Together, these works suggest a shift from single-skill tracking toward reusable motion priors and scalable physics-based controllers.

Another related direction studies competitive and multi-agent humanoid behavior. In such settings, the agent must not only imitate individual motions, but also adapt to another agent whose behavior changes over time. Deep RL has been used to train humanoid robots for dynamic soccer, where learned locomotion, kicking, recovery, and self-play strategies are composed into competitive behavior \cite{haarnoja2024learning}. For contact-rich combat, multi-agent adversarial imitation and self-play methods have been explored to model interactive fighting behaviors. For example, MAAIP extends adversarial imitation to multi-character fighting demonstrations, while RoboStriker uses a hierarchical framework with motion tracking, latent skill representation, and latent-space neural fictitious self-play to learn autonomous humanoid boxing tactics \cite{younes2023maaip,yin2026robostriker}. KungfuBot further demonstrates the feasibility of physics-based whole-body control for highly dynamic martial arts-style motions using motion processing, adaptive tracking, and RL \cite{xie2026kungfubot}. 

Across these areas, evaluation largely emphasizes controller-oriented metrics such as tracking error, task success, stability, robustness, and energy efficiency, supplemented by qualitative demonstrations \cite{peng2018deepmimic,peng2021amp,peng2022ase,escontrela2022adversarial,haarnoja2024learning}. Although these measures establish functional and physical performance, they do not capture how domain practitioners or users perceive motion plausibility, fluidity, and stylistic authenticity in application-oriented settings. This gap motivates both our user evaluation of the learned motions and the development of the proof-of-concept VR sparring game prototype.

\subsection{Interactive Agents in VR Combat Games}
Traditional combat game agents typically couple scripted decision logic with pre-authored animation assets: rules, finite-state machines, behavior trees, or planning modules map the player’s state or input to a corresponding animation or procedural response. Although predictable and designer-controllable, this approach limits expressiveness to predefined behaviors and transitions and becomes increasingly difficult to author as the number of actions and interaction contexts grows.

Recent work on VR combat has explored responsive opponent behavior, personalized training, and embodied feedback. Touché recognizes player actions in VR sword fighting and synthesizes responsive opponent motions from recorded combat examples \cite{dehesa2020touche}, while KaraKter combines motion-captured techniques with decision logic to select attacks against a tracked athlete \cite{zhang2018karakter}.

More directly related to VR martial arts training, Takala et al. developed a full-body system in which a tracked user avatar and virtual opponents interact through physics simulation \cite{takala2020martial}. This allows punches, kicks, takedowns, and grappling actions to physically affect opponents. The system also supports sparring and target practice, adjustable opponent properties, and replay of user-recorded techniques through a virtual opponent. VCoach \cite{chen2022vcoach} generates personalized boxing drills from real-time measures of punch speed, reaction time, and pose, presenting target sequences and feedback on a virtual heavy bag. Tian et al. developed Kung Fu Metaverse, a movement-guidance system containing nearly one hundred traditional Chinese Kung Fu routines \cite{tian2023kung}. 

Other systems emphasize physical contact. MovableBag uses a room-scale mobile substitutional robot capable of carrying sports props to simulate a boxing opponent or coach with encountered-type haptic feedback \cite{mendez2022movablebag}. Hit Around further supports mobile combat interaction by allowing users to move around and strike a virtual opponent while receiving encountered-type haptic contact \cite{weng2025hit}. Related gesture-to-action and generative-agent methods \cite{clark2016system,kang2020deephandsvr,gunawardhana2024toward} also improve responsiveness to tracked users. Robo-Saber generates style-conditioned headset and controller trajectories for automated Beat Saber playtesting and converts them into whole-body motion through physics-based tracking \cite{kim2026robo}. Together, these studies advance responsive interaction, movement guidance, haptic feedback, and VR player simulation; however, learning reusable physics-based martial arts skills for responsive opponents remains comparatively underexplored.

\section{Motion Data Collection}
Tang Soo Do is a traditional Korean martial art influenced by Shotokan Karate and Chinese martial arts. We selected it because its emphasis on foundational body mechanics and its diverse repertoire of linear and circular attacks, defenses, and transitions make it well suited to humanoid behavior learning and virtual sparring. We collaborated with a local Tang Soo Do association for motion capture.
\begin{figure}
  \centering
  \includegraphics[width=\linewidth]{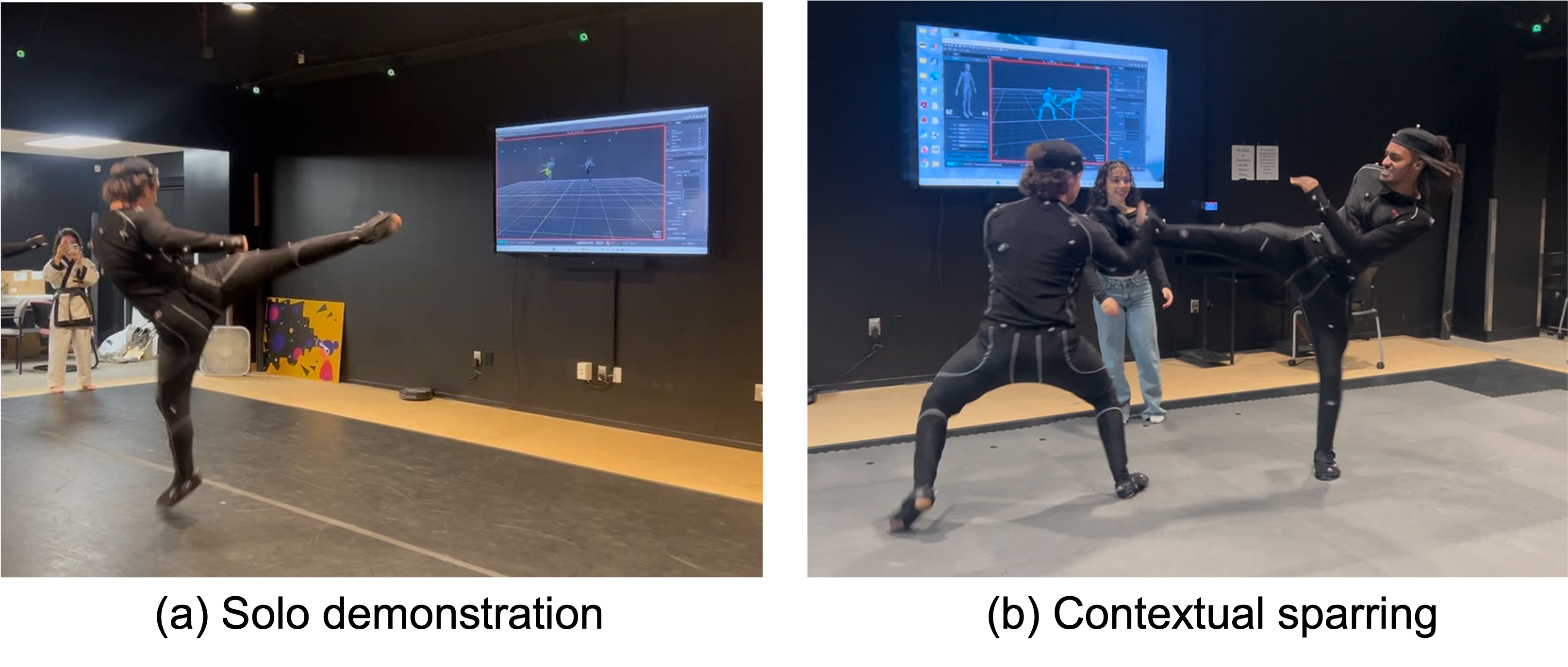}
  \caption{Motion-capture settings used across the staged martial arts data-collection protocol.}
  \label{fig:motion_capture}
\end{figure}
\subsection{Participants}
We recruited five experienced Tang Soo Do practitioners (one female and four male practitioners) from the association. All held the rank of 1st Dan (black belt) or higher and had more than five years of experience in Tang Soo Do and martial arts more broadly. Their expertise supported technically proficient executions while preserving natural variation in individual movement styles. All human-participant procedures described in this paper were approved by our institutional ethics review board before data collection, and all participants provided informed consent.
\subsection{Motion Capture Configuration}
Motion data were recorded using an OptiTrack motion capture system installed in our department's motion capture studio. The capture volume was calibrated before recording to support accurate tracking of the rapid and large-amplitude movements commonly observed in martial arts techniques.

Participants were equipped with a standard retro-reflective marker set, and the tracked marker trajectories were used to reconstruct skeletal motion. The OptiTrack system operated at 120 Hz during capture. For subsequent processing and training, the motion sequences were represented at 30 FPS to improve computational efficiency.
\subsection{Movement Protocol}
To capture both the technical repertoire of Tang Soo Do and the contextual variations that emerge during interaction, we designed a three-phase motion acquisition protocol:
\begin{enumerate}
    \item \textbf{Isolated Techniques:} One or more participants performed repeated trials of movements they regularly practiced, including basic punches, kicks, and blocks, as well as more advanced skills. This phase establishes a motion dictionary of practitioners’ technical repertoire.
    \item \textbf{Short Combinations:}
     One or more participants performed repeated trials of short sequences that combined punches, kicks, blocks, and transitions. This phase captures transitional dynamics, temporal coordination, and weight shifting between individual techniques.
    \item \textbf{Two-Person Sparring:}
    Pairs of participants engaged in simulated sparring without predetermined action sequences. Participants naturally responded to their partners with attacks, defensive responses, footwork, and changes in distance. This phase captures reactive and context-dependent variations of martial arts movements.
\end{enumerate}

Across all recording sessions, we collected 11.2 minutes of wall-clock capture time and 17.8 minutes of aggregate motion data, with the latter summing the individual motion tracks of all participants in multi-person recordings. Detailed information on motion data retained for training after preprocessing and trimming is provided in Section \ref{sec: experiments}.

The three-phase protocol provides complementary motion content. Technically demanding rotational and acrobatic movements, including spinning and jumping kicks and flips, were captured during isolated demonstrations rather than free sparring, where such skills were less commonly used. In contrast, the combination and sparring recordings captured transitional behaviors such as dynamic combat idle, weight shifting, guard adjustment, and preparatory chambering. Retaining both isolated and interaction-based recordings therefore provided discrete skills, transitions, and reactive behaviors for subsequent humanoid training and virtual sparring.

\section{Humanoid Skill Learning}
Motion style encompasses the behavioral characteristics that define the overall ``feel'' of movement. Although humans intuitively perceive these nuances, they are difficult to encode in computational objectives. Conventional motion imitation often emphasizes pose-tracking accuracy, while task-oriented reinforcement learning prioritizes objective completion; both may fail to preserve natural fluidity and style. AMP \cite{peng2021amp} addresses this issue by combining reinforcement learning with an adversarial motion prior, enabling a task-specific policy to optimize both task and style rewards. Building on this adversarial formulation, ASE \cite{peng2022ase} introduces latent skill variables and a diversity objective to learn a reusable skill space from a large and diverse motion dataset. We extend this foundation to martial arts combat by pretraining a latent repertoire of martial arts skills and learning task-specific controllers with customized rewards for target-directed striking and threat-aware evasion. This enables a shared skill repertoire to be selected and adapted in response to changing targets and incoming threats, producing goal-directed behaviors that remain grounded in the motion characteristics of the captured martial arts demonstrations.

\subsection{Problem Formulation and Pretraining}
In physics-based humanoid control, the learning problem is formulated as learning a policy that maps the current humanoid state to a control action. At each simulation step $t$, the policy observes the current state $s_{t}$, which is determined by the input features of the articulated skeleton, including the relative position of each body link and the rotation of each link. The action $a_{t}$ is the output of the policy, specifying the target rotations for PD controllers positioned at each of the humanoid’s joints. These targets are then passed to the virtual character to move the joints toward the desired configuration.

To train the policy, we follow the dual-reward formulation of AMP and optimize the agent with both task-driven and style-driven objectives. Formally, the policy $\pi_\theta(a_t \mid s_t)$ is trained to maximize the expected discounted return:
\begin{equation}
J(\theta)=\mathbb{E}{\pi\theta}\left[
\sum_{t=0}^{T}\gamma^t
\left(
w_{\text{task}} r_t^{\text{task}} +
w_{\text{style}} r_t^{\text{style}}
\right)
\right],
\end{equation}
where the reward consists of two parts: a task reward and a style reward. The task reward $r_t^{\text{task}}$ is handcrafted, specifying the high-level objectives. 
The style reward $r_t^{\text{style}}$ is generated by an adversarial discriminator and reflects how likely the discriminator is to classify a generated motion transition as originating from the reference data.
This dual-reward structure enables the agent to perform specific tasks while maintaining certain motion styles.

During the pretraining stage, the agent is not optimized for specific combat tasks. Instead, following ASE, the pretraining objective combines an adversarial style reward with a skill-discovery reward:
\begin{equation}
r_t^{\text{pretrain}} = r_t^{\text{style}} + \beta r_t^{\text{skill}}.
\end{equation}
Here, $\beta = 0.5$. The discriminator provides an adversarial motion prior that encourages generated motion transitions to match the reference motion dataset, while the encoder encourages the generated behaviors to preserve information about the sampled latent variable and maintain diversity across latent-conditioned skills. The collected trajectories are then used to update the discriminator, encoder, and policy. The policy is optimized using proximal policy optimization (PPO) \cite{schulman2017proximal}.

After pretraining, the controller has learned a reusable motion prior, i.e., a diverse repertoire of martial-arts movements. However, these motions are not yet conditioned on specific combat objectives. We therefore introduce task-specific training, which learns to adapt the pretrained motion skills to different task conditions by selecting latent commands according to the current target.

\subsection{Task-Specific Training}
\subsubsection{Target-Directed Striking}
\begin{figure}
  \centering
  \includegraphics[width=\linewidth]{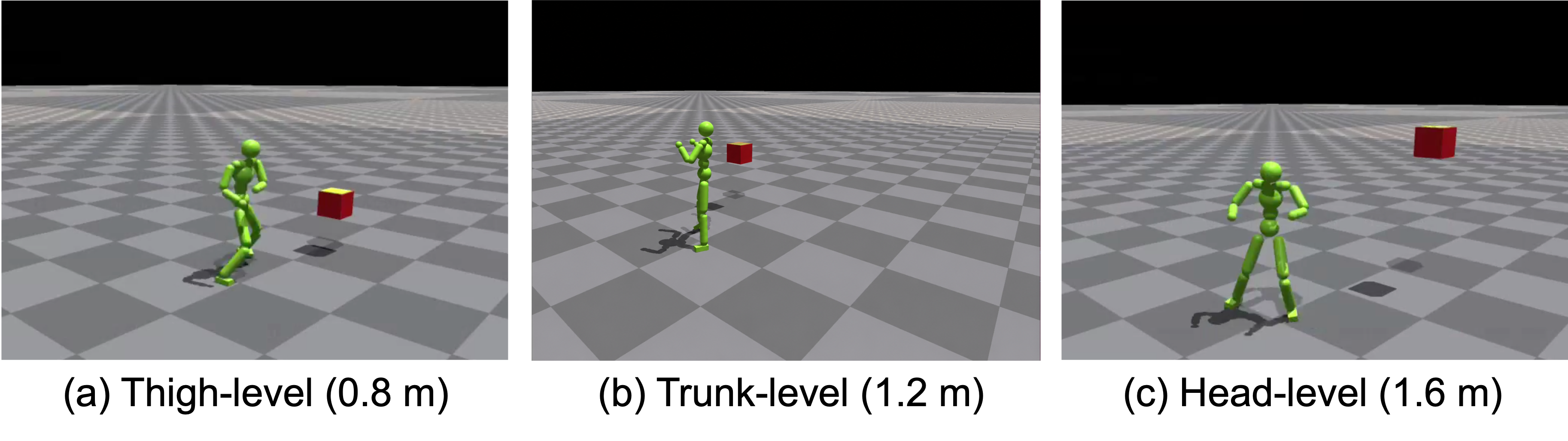}
  \caption{Target configurations used for the striking training.}
  \label{fig:striking}
\end{figure}

In real-life martial arts training, students learn to strike a heavy bag or pad before practicing against a moving opponent. To mimic this process, we design a task-specific reward to enable the humanoid to strike a virtual object using hands or feet while maintaining consistency with the learned martial arts style. To simulate different target regions of an opponent's body, we generate a cube at random horizontal positions within a predefined range around the humanoid and at one of three predefined heights: 0.8 m, 1.2 m, and 1.6 m, corresponding approximately to thigh-, trunk-, and head-level interactions, respectively (Fig. \ref{fig:striking}). The trunk and head levels represent the common target areas in sparring, while the thigh-level target is motivated by our observations that practitioners also perform low foot or hand movements around this height to intercept or block an opponent's incoming attack, even when the thigh itself is not a primary scoring area. The cube's location changes every 5 s, and once the humanoid successfully strikes the cube, it disappears and reappears at a new random location. This design encourages the agent to adapt its learned striking motions to different target positions and heights rather than repeatedly executing a single fixed strike.

Following AMP/ASE, the total reward combines the task reward with an adversarial style reward $r_t^{\text{style}}$ from a discriminator trained using the reference motion dataset: 
\begin{equation}
r_t^{\text{striking}} = w^{\text{task}}r_t^{\text{task}} + w^{\text{style}}r_t^{\text{style}}.
\end{equation}
Here, $w^{\text{task}} = 0.9$ and $w^{\text{style}} = 0.1$. This formulation encourages the controller to accomplish the striking objective while selecting behaviors that remain consistent with the martial-arts motions learned during pretraining.

We design the task reward to distinguish an effective strike from incidental contact. Specifically, a successful strike should produce a sufficiently large change in the cube's orientation and satisfy a minimum impact requirement. The task reward is a weighted sum of an orientation term that measures how far the pad has been knocked from upright, and a velocity term that shapes the approach:
\begin{equation}
r_t^{\text{task}} = w^{\text{rot}}r_t^{\text{rot}} + w^{\text{vel}}r_t^{\text{vel}}.
\end{equation}
We set $w^{\text{rot}} = 0.6$ and $w^{\text{vel}} = 0.4$. The orientation and velocity terms are applied only when their respective thresholds are met; otherwise, the terms will be set to zero. These constraints discourage behaviors in which the hand or foot merely grazes the target without delivering a meaningful strike. Only designated striking body parts, including the hands and feet, are considered valid for task completion.

\subsubsection{Threat-Aware Evasion}
In Tang Soo Do, defensive responses can take many forms. For the purpose of our task design, we broadly group them into evasion and blocking/interception. Evasion avoids an incoming attack by changing the defender's distance, angle, or body position, and is commonly achieved through footwork such as backward steps, lateral steps, diagonal steps, and pivoting or turning movements. Blocking, in contrast, uses the limbs to cover, redirect, or absorb an incoming strike, with common examples including upper- and middle-level blocks that protect the head and torso. Defensive responses may also involve interceptive actions, such as checks, stop-kicks, or counter-strikes that disrupt an opponent's attack before it fully develops. In actual sparring, these defensive behaviors are often combined rather than performed as isolated actions, making defense highly dynamic and context dependent.
\begin{figure}
  \centering
  \includegraphics[width=\linewidth]{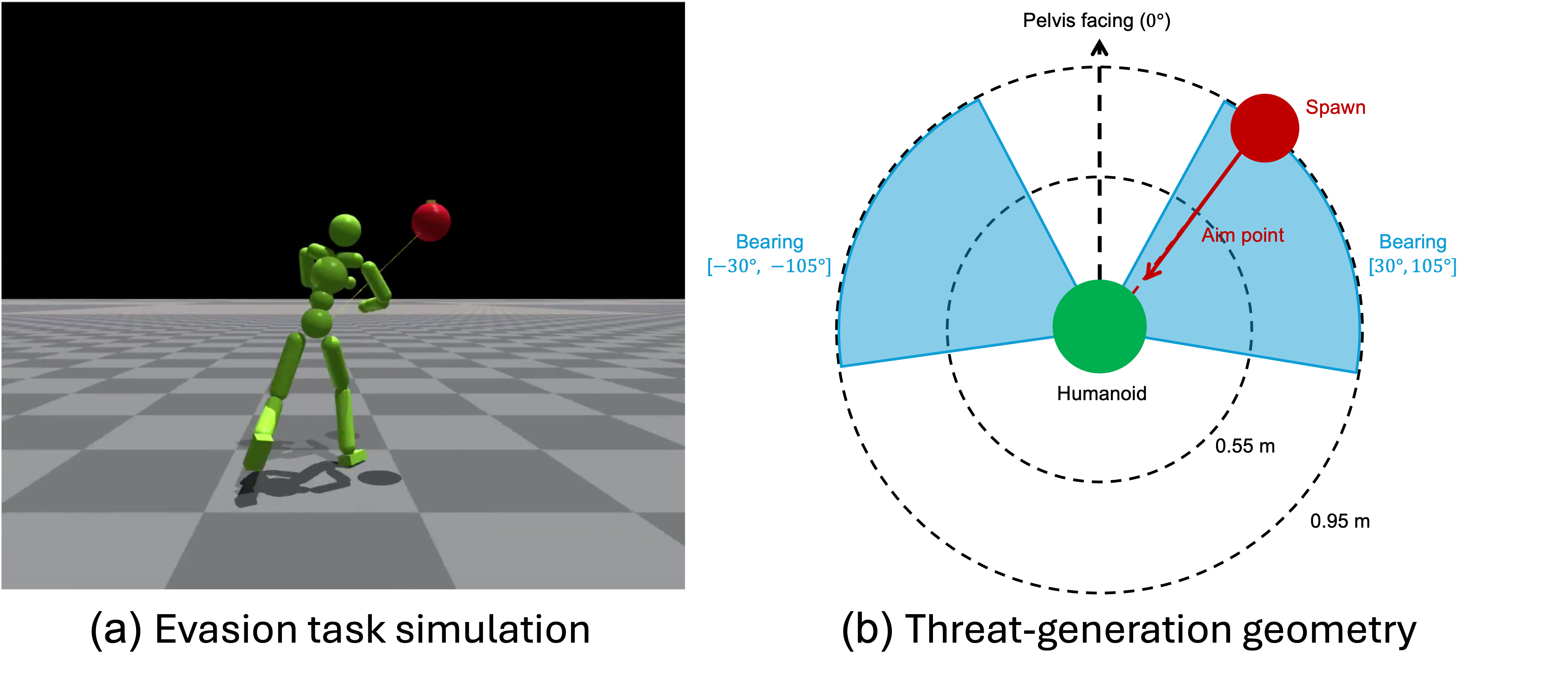}
  \caption{Threat-generation setup for evasion training.}
  \label{fig:evasion}
\end{figure}
The pretrained motion repertoire already includes several representative stationary blocking techniques. Therefore, in this task-specific training stage, we focus primarily on defensive responses that involve spatial repositioning, including evasive footwork and movement-based defensive actions. The objective is to encourage the humanoid to reposition itself in response to an approaching threat while remaining within a realistic fighting distance.

As shown in Figure \ref{fig:evasion}, we model an incoming strike as a sphere with a radius of 0.12~m. At the beginning of each episode, the sphere is spawned at a random location within the blue regions shown in Figure \ref{fig:evasion}(b). Its bearing, measured relative to the humanoid's pelvis-facing direction, is sampled from $[-105^\circ,-30^\circ]\cup[30^\circ,105^\circ]$, and its horizontal distance from the humanoid ranges from 0.55~m to 0.95~m. These bounds were estimated from the spatial distributions of incoming strikes in evasion clips extracted from the captured sparring sessions. Absolute bearing angles below $30^\circ$ occurred infrequently in these recordings and were therefore excluded from the spawn distribution. Sampling within the observed ranges produces simulated threats consistent with the spatial configurations found in real sparring interactions.

The evasion task follows the same task--style reward formulation used for target-directed striking. The style reward is provided by the discriminator trained during skill pretraining, while the task reward consists of clearance and engagement components:
\begin{equation}
r_t^{\text{task}}
= w^{\text{clear}} r_t^{\text{clear}}
+ w^{\text{engage}} r_t^{\text{engage}} .
\end{equation}
We set $w^{\text{clear}}=0.7$ and $w^{\text{engage}}=0.3$. The clearance term, $r_t^{\text{clear}}$, encourages the humanoid to move away from the predicted path of the incoming strike.
Instead of considering only the sphere's current position, we measure clearance from its anticipated trajectory. This provides an earlier learning signal, allowing the controller to begin an evasive response before the strike reaches the humanoid. The reward saturates once a sufficient clearance is reached so that the policy is encouraged to perform a controlled evasion rather than simply move as far away as possible.

The engage term, $r_t^{\mathrm{engage}}$, discourages excessive retreat. The humanoid may reposition freely within a small region around its location at the beginning of the strike, while larger displacement receives progressively lower reward. This encourages defensive movements that avoid the attack while keeping the humanoid within a realistic sparring distance.

Finally, when each attack ends, we additionally evaluate the minimum physical clearance between the incoming sphere and the humanoid during the event. This provides a direct success signal based on whether the sphere actually reached the humanoid, complementing the predictive clearance reward used during the approach. Together, these rewards encourage the controller to select evasive movements that anticipate an incoming attack and avoid contact.

\section{Evaluation}
\subsection{Simulation-Based Policy Evaluation}
\label{sec: experiments}
\subsubsection{Experimental Setup}
\begin{figure}
  \centering
  \includegraphics[width=\linewidth]{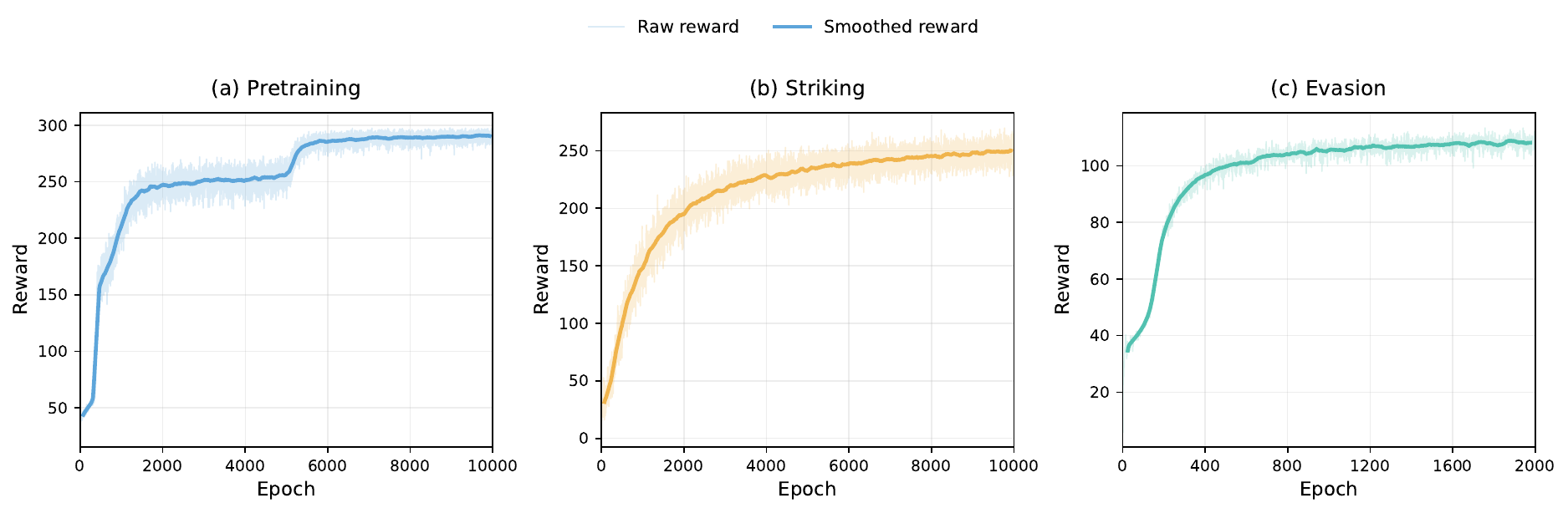}
  \caption{Training reward curves for (a) pretraining, (b) striking, and (c) evasion.}
  \label{fig:reward}
\end{figure}

\begin{figure}
  \centering
  \includegraphics[width=\linewidth]{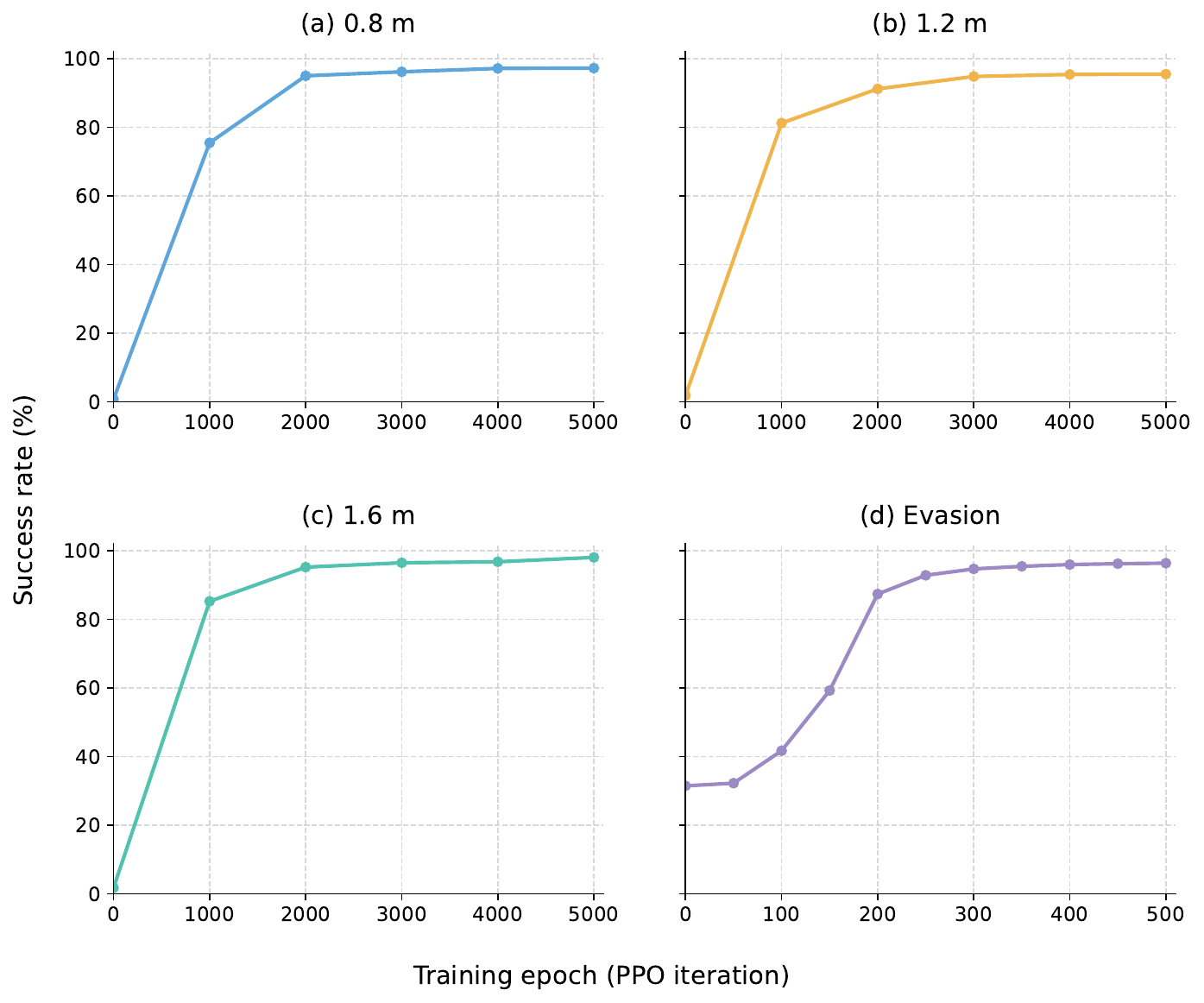}
  \caption{Task success rates over training for (a)-(c) striking at each pad height and (d) evasion.}
  \label{fig:strike_success_rate}
\end{figure}

All experiments were conducted in NVIDIA Isaac Gym with 4,096 parallel environments. The simulator runs at 60 Hz, while the control policy operates at 30 Hz. Each episode lasts 300 control steps, corresponding to 10 s. The humanoid has 15 rigid bodies, 28 actuated degrees of freedom (DoFs), and a free root, and is controlled using PD targets.

After preprocessing, including trimming, the dataset retained for pretraining consisted of 97 original motion sequences: 26 non-sparring sequences (330.0 s), 41 attacking motion segments extracted from the sparring sequences, referred to as sparring-attack sequences (58.1 s), and 30 evasive motion segments extracted from the sparring sequences, referred to as sparring-evasion sequences (26.7 s). Together, these sequences comprise approximately 414.8 s (6.9 min) of motion data. To increase the amount of training data and reduce directional bias, we additionally generated a mirrored version of each retained sequence. This augmentation doubled the training set to 194 motion sequences, corresponding to approximately 829.6 s (13.8 min) of motion data.

For target-directed striking training, we constructed a task-specific subset by selecting attacking motions from the non-sparring sequences and including the sparring-attack sequences. For threat-aware evasion training, we used only sparring-evasion sequences.
\subsubsection{Quantitative Results} 
\label{sec: quantitative_results}
The reward curves in Figure~\ref{fig:reward} reveal distinct learning dynamics across the three stages, although their absolute values are not directly comparable because each uses a different objective. Pretraining exhibits a two-stage pattern: the reward rises rapidly during the first 2,000 epochs, plateaus near 250, and then increases again around 5,000--6,000 epochs before stabilizing near 290. This pattern may reflect the policy learning common motion patterns first and improving less frequent or more complex skills later. Striking improves steadily throughout training, whereas evasion approaches its plateau within approximately 600 epochs. The rapid convergence of evasion training is consistent with both its narrower task formulation and the limited coverage of defensive motions in the dataset, which were extracted from free sparring rather than intentionally collected. Therefore, it reflects efficient learning within the current threat distribution but not necessarily a comprehensive defensive repertoire, motivating a more targeted collection of evasive behaviors in future work.

The success rate curves in Figure~\ref{fig:strike_success_rate} show that basic task competence emerges before the shaped rewards fully converge. All three striking conditions exceed 90\% success after approximately 2,000 epochs and ultimately reach about 97\%, 95\%, and 98\% for the 0.8~m, 1.2~m, and 1.6~m targets, respectively. Although the thigh-height target is learned more slowly initially, this difference largely disappears with continued training. Evasion success rises from approximately 30\% to nearly 90\% by 200 epochs and approaches 97\% after 300 epochs. The continued reward improvement after success rate begins to stabilize suggests that later training refines target alignment, clearance, engagement, and stylistic consistency rather than merely increasing task completion.

\subsubsection{Qualitative Motion Analysis}
\begin{figure*}
  \centering
  \includegraphics[width=0.8\linewidth]{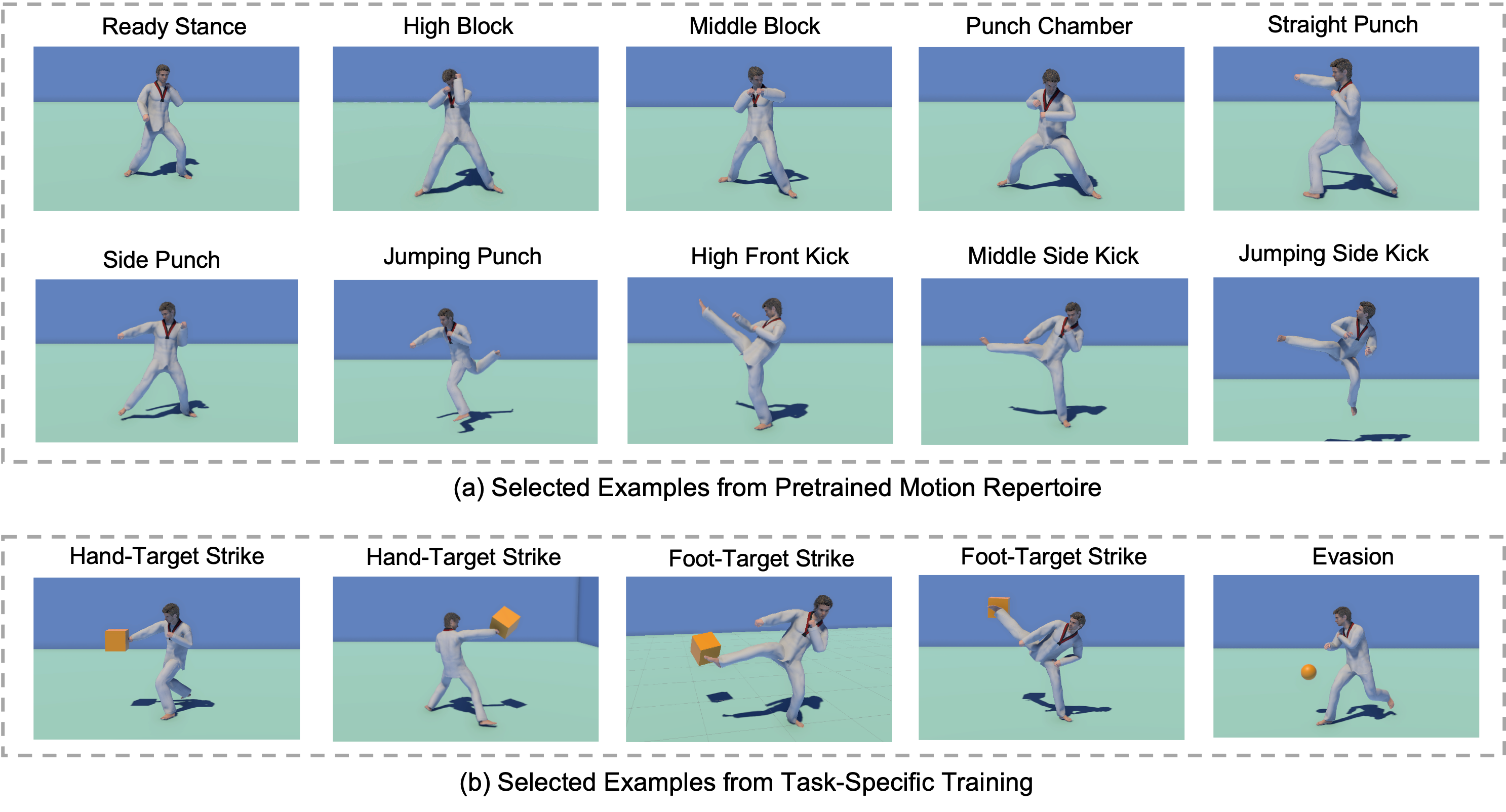}
  \caption{Selected examples of motions generated by the learned policies.}
  \label{fig:generated_motions}
\end{figure*}
Figure~\ref{fig:generated_motions} presents representative examples of the generated martial arts motions. For clearer visualization, trajectories produced by the physics-based humanoid policies were retargeted to a humanoid avatar in Unity. As shown in Figure~\ref{fig:generated_motions}(a), the pretrained repertoire spans foundational movements, including ready stances, punches, kicks, and blocks, as well as more complex jumping techniques that require coordinated balance and whole-body control. The generated sequences also include preparatory and transitional movements, such as punch chambering, rather than reproducing only the most visually salient attack or defensive poses. This indicates that the policy captures the broader temporal structure of each technique.

Figure~\ref{fig:generated_motions}(b) shows how the task-specific controllers adapt this repertoire to changing task conditions. For striking, the target cube is randomized in height and distance, requiring the controller to adjust limb elevation, body extension, and stance to reach the target with either the hand or foot. These variations show that the controller modifies learned skills according to the target location rather than replaying fixed motion clips. For evasion, the controller combines body repositioning with a guarded posture in response to the incoming threat, balancing task-driven avoidance with the martial arts style acquired during pretraining. Together, these examples illustrate how the pretrained skill space supports both a diverse motion repertoire and goal-directed adaptation. Detailed user ratings and qualitative feedback are presented in the next section.

\subsection{Practitioner User Study}
\begin{figure*}
  \centering
  \includegraphics[width=\linewidth]{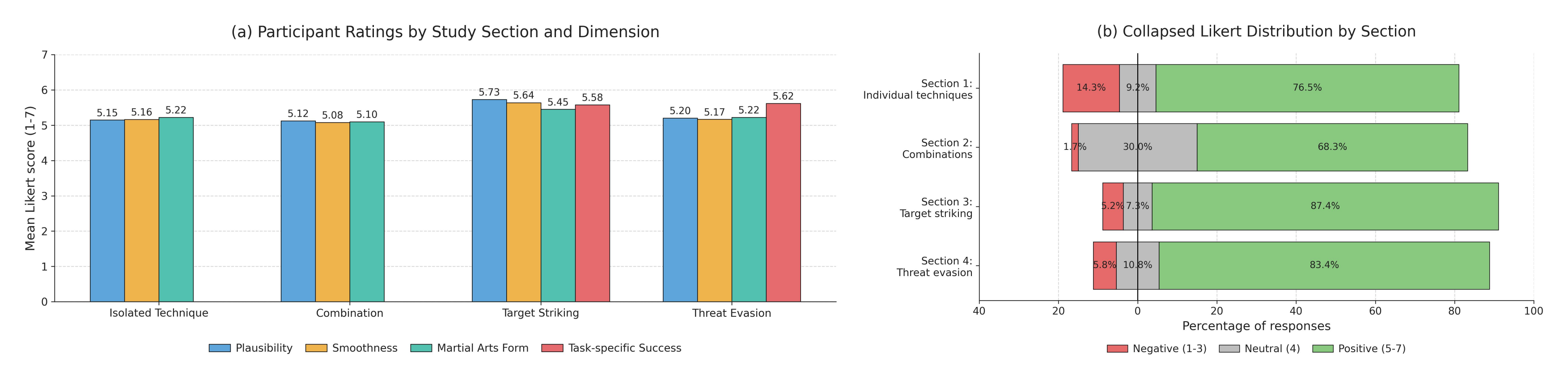}
  \caption{Motion quality evaluation across study sections.}
  \label{fig:motion_quality}
\end{figure*}
\begin{figure*}
  \centering
  \includegraphics[width=\linewidth]{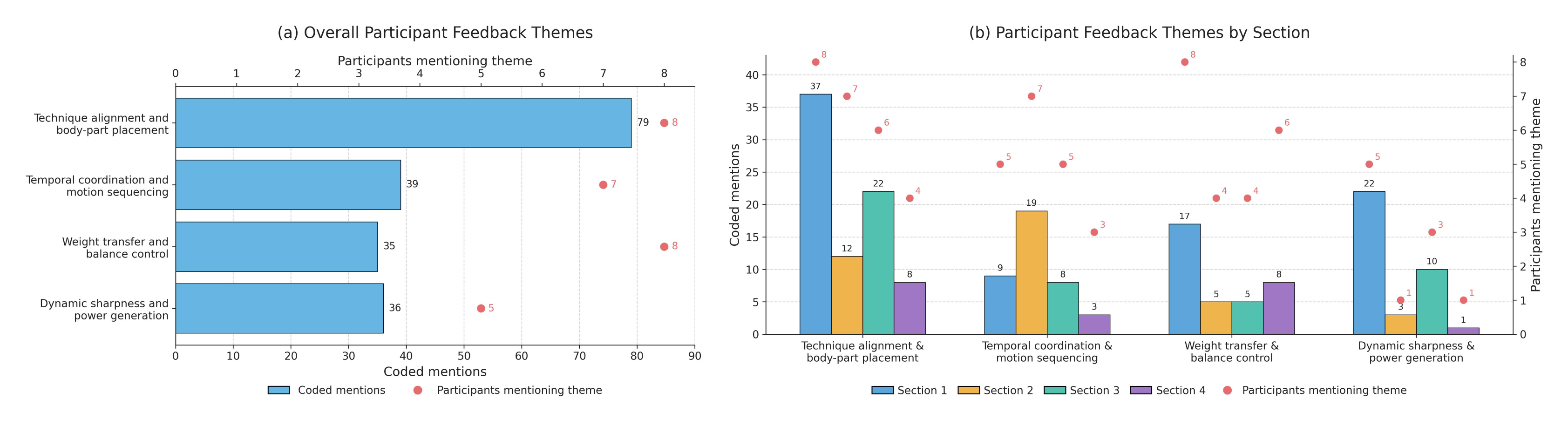}
  \caption{Qualitative feedback themes in the open-ended questions.}
  \label{fig:feedback_theme}
\end{figure*}
To further evaluate the generated motions, we conducted a user study in which domain practitioners assessed motion quality from martial arts and game-development perspectives. The central questions were: 
\begin{enumerate}
    \item Can the framework generate plausible, realistic, and smooth motions sufficient to support an engaging user experience?
    \item Do these motions successfully complete intended tasks while preserving an authentic martial arts style across scenarios?
\end{enumerate}
To achieve these goals, we built a Unity-based GUI and retargeted the humanoid motions generated in Isaac Gym onto a game avatar. This allowed participants to evaluate the motions in a more application-oriented setting.




\subsubsection{Participants and Procedures}
Participants are recruited through advertisements distributed to local martial arts associations and the university's student martial arts clubs. In total, eight participants took part in the user study. Their ages ranged from 20 to 40. Six were male and two were female. They all have at least five years of martial arts experience and prior experience with fighting or martial arts video games.

After providing informed consent, participants were assigned anonymous user IDs. Each in-person session began with an overview of the project and study procedure, followed by a calibration session to familiarize participants with the evaluation criteria. Participants viewed examples of human-authored martial arts animations and motions directly retargeted from the captured data. The formal evaluation began after they confirmed their understanding of the task, criteria, and reference examples.

The study consisted of four sections. Section 1 included ten clips showing isolated martial arts techniques, covering fundamental blocks, punches, and kicks. Section 2 included five clips randomly sampled from the model’s latent space. These clips represented short combinations of multiple techniques. Section 3 included eight target-striking clips: four hand strikes and four foot strikes. Section 4 included four clips of threat-evasion tasks. Except for the randomly sampled combination clips in Section 2, clips were manually selected to ensure coverage of the predefined technique categories. To limit selection bias, manual selection was based on category coverage and sequence completeness rather than perceived motion quality.

Participants viewed each motion clip in the 3D Unity scene and evaluated it using a seven-point Likert scale. A score of 1 corresponded to ``Strongly disagree'', and a score of 7 corresponded to ``Strongly agree''. For all clips, they rated each motion on three dimensions: whether the character’s movement appeared physically plausible, balanced, and human-like; whether the motion was fluid and free of unnatural jitters, stiffness, or abrupt transitions; and whether the movement reflected proper martial arts form, technique, and style. For the target-striking clips in Section 3, participants rated one additional aspect by evaluating whether the character convincingly struck the target with appropriate reach and alignment. For the evasion clips in Section 4, participants rated one additional aspect by evaluating whether the character convincingly evaded the threat with appropriate distance and footwork. Throughout all sections, participants were encouraged to provide written comments explaining their ratings.

\subsubsection{Results and Insights}
As shown in Figure \ref{fig:motion_quality}(a), ratings generally leaned toward agreement across all four sections. Overall ratings were highest for the target striking motions (M = 5.61, SD = 1.10), followed by the threat evasion motions (M = 5.30, SD = 1.15). On the seven-point scale, these values fell between ``Slightly agree'' and ``Agree'', indicating generally favorable but not uniformly strong assessments. Ratings of plausibility, smoothness, and martial arts form followed a similar pattern, suggesting that participants tended to perceive the generated motions as believable. Figure \ref{fig:motion_quality}(b) provides a more detailed view of the variation in these responses.

This moderate tendency toward agreement is especially informative when compared with the quantitative task results. As reported in Section \ref{sec: quantitative_results}, the success rates for both striking and evasion exceeded 95\%, whereas the corresponding mean task-specific Likert-scale ratings were 5.58 and 5.62, respectively. This difference reveals an important distinction between task completion and perceived technical correctness. The automated criteria measured whether a strike produced sufficient target impact or an incoming threat was avoided, whereas participants also judged the technical execution of the motion. Consequently, a strike could satisfy the quantitative criterion while still appearing technically incorrect because it used an unintended body part or contact surface, or exhibited suboptimal alignment, weight transfer, or timing. Similarly, an evasion could successfully avoid the threat while using less appropriate footwork or weight transfer.

In addition to the rating results, we further examined the participants' open-ended comments to better understand the factors behind their evaluations. Rather than indicating that all generated motions suffered from the same flaws, these comments revealed several localized failure modes that appeared in specific clips, especially those receiving relatively lower ratings. We summarized the observed technical flaws under four categories:
\begin{enumerate}
    \item \textbf{Technique refinement:} 
    specific aspects of some motions did not achieve the technical precision expected by experienced practitioners.
    \item \textbf{Weight transfer and balance control:} 
    the motion's weight distribution did not match common martial arts practice.
    \item \textbf{Dynamic sharpness and power generation:} 
    lacked sufficient acceleration, crispness, or a sharp endpoint.
    \item \textbf{Temporal coordination and motion sequencing:}
    in combinations, the techniques were not connected with appropriate timing.
\end{enumerate}

Figure \ref{fig:feedback_theme}(a) and Figure \ref{fig:feedback_theme}(b) show that the refinement themes were not evenly distributed across motion types. Technique refinement was the most common theme, especially in isolated techniques and target-striking clips, where pose accuracy and striking endpoints were easier to inspect. One participant noted that ``the punch lacks strong hip rotation, shoulder commitment, and a clean retraction of the opposite hand,'' suggesting that the motion was recognizable but not technically precise. Temporal coordination was concentrated in combinations, suggesting that chaining techniques with convincing rhythm and timing remains challenging. Another participant commented that the techniques ``do not connect with the sharp timing and structure expected in an ideal kick-punch-block sequence.'' Weight distribution issues appeared more locally in a few blocking or kicking motions, where the motions were physically balanced but less consistent with martial arts practice. Several participants also commented on weight transfer in the threat-evasion clips, noting that the weight shifts accompanying some steps were less pronounced than expected. For dynamic sharpness and power generation, participants also referred to “motion crispness” or “motion acceleration”, which we grouped under the same theme. Because martial arts techniques are expected to be fast, forceful and decisive, sluggish motion can noticeably weaken the user experience. One participant noted that handcrafted combat animations often use exaggerated acceleration, snap, and impact emphasis, suggesting that these qualities could be further refined in our generated motions. 

Overall, the study suggests that the learned policy provides a promising foundation for generating virtual sparring motions. More broadly, the findings distinguish task completion from technique correctness: contact-based success criteria should account for the intended body part and contact surface, while motion priors should represent the complete temporal and mechanical structure of a technique, including preparation, alignment, weight transfer, acceleration, impact timing, and recovery, rather than only its salient attack pose.
\section{VR Sparring Game Prototype}
\begin{figure}
  \centering
  \includegraphics[width=\linewidth]{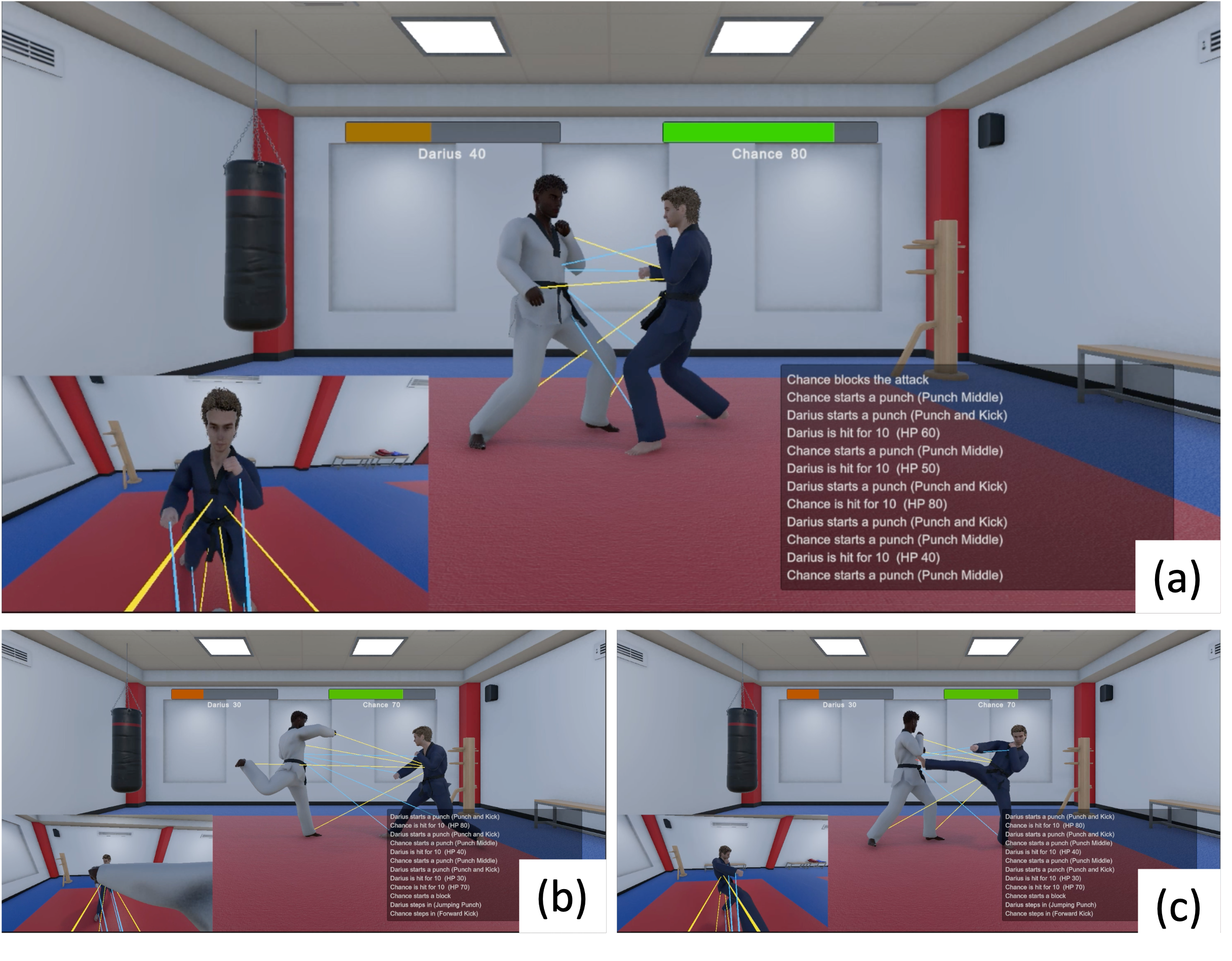}
  \caption{Representative interactions from the prototype: (a) Both avatars in an idle state; (b) Darius starts a jumping punch while Chance evades with backward steps; (c) Chance steps forward with a kick to close the distance to Darius.}
  \label{fig:game_scene}
\end{figure}

To examine how the learned motions could be deployed in an interactive application, we developed a proof-of-concept VR sparring game in Unity for Meta Quest headsets. Figure~\ref{fig:game_scene} presents representative interactions from the prototype. The avatar on the left, Darius, represents the human player, while the avatar on the right, Chance, serves as the virtual opponent. The prototype does not execute the learned policies online. Instead, policy-generated rollouts were segmented into action clips, retargeted from the simulated humanoid to the Unity character skeleton, and associated with gameplay metadata such as action states and active contact windows.

At runtime, both characters execute retargeted clips, but their actions are selected through different control pathways. Predefined movement thresholds map the player's controller gestures to Darius's action states, which trigger corresponding motions from the learned skill library. Chance is governed by lightweight state-based opponent logic that monitors the distance between the avatars' trunks and end effectors. When one of Darius's end effectors approaches Chance, the system identifies a potential attack and selects an appropriate block or evasion. If no threat is detected within a predefined time window, Chance initiates an attack. Because evasive actions may increase the distance between the avatars, both characters can execute actions containing forward steps to reestablish an effective sparring distance. An optional distance-monitor visualization displays the monitored distances and highlights detected threats, helping the player anticipate the opponent's responses.

Contact events are detected using colliders attached to the avatars' hands, feet, and trunks together with action-specific active windows. An overlap during an attack's active window registers a single hit, while a correctly timed block prevents damage. Because the prototype does not provide physical impact or haptic feedback, successful attacks, blocks, and other recognized interactions are communicated through impact sounds, health-bar changes, and event logs.

This hybrid architecture provides a practical integration path for deploying learned motions in a consumer VR application. However, converting policy rollouts into animation clips means that the opponent adapts only through discrete state transitions rather than through continuous physics-based control. The prototype therefore demonstrates the applicability of the learned motion repertoire to interactive sparring, while a fully adaptive, physics-based virtual opponent remains future work.

\section{Limitations and Future Work}
We present this prototype as an initial step toward a more comprehensive interactive system. To prioritize the core functionality for this proof-of-concept, we deliberately scoped peripheral components with lightweight implementations. These design trade-offs, alongside insights from our expert user study, present clear pathways for future research.

First, the current prototype relies on a hybrid pipeline with lightweight state-transition logic to organize generated motion clips. While sufficient for demonstrating martial arts skill integration, it lacks a fully autonomous sparring opponent. Future iterations will replace this pipeline with an intelligent system capable of real-time, adaptive combat decision-making in response to player actions.

Second, player input currently utilizes controller gestures to trigger retargeted motions, which limits physical immersion. To evolve this system into a fully embodied game, future work will enable players to control their avatars through full-body input, substantially enhancing the sense of embodiment.

Finally, while our initial evaluation focused on the perceived quality of the generated motions, future studies will evaluate real-time interactive gameplay. Having users actively play against the virtual opponent will allow us to measure broader experiential and training factors, such as system responsiveness, user engagement, perceived opponent intelligence, and training efficacy. This will provide a holistic understanding of how learned humanoid policies impact virtual martial arts practice.
\section{Conclusion}
This paper presented a skill-based motion learning framework for virtual martial arts opponents. Using a three-phase Tang Soo Do motion dataset, we trained a physics-based humanoid with adversarial skill learning and task-specific rewards for striking and evasion. The generated motions were generally perceived as physically plausible, smooth, and recognizable as martial arts techniques, although further improvements are needed in technique precision, weight transfer, dynamic sharpness, and temporal coordination. We also integrated the motions into a virtual sparring game prototype runnable on Meta Quest headsets, demonstrating their potential to support interaction between a human player and a state-controlled opponent. Future work will develop a fully autonomous opponent, improve player input, and evaluate the system through interactive user studies.


\bibliographystyle{abbrv-doi}

\bibliography{template}
\end{document}